\documentclass[12pt]{article} 
\usepackage{graphicx}
\usepackage{amsmath,amssymb,amsfonts,longtable}

\begin{document}
\baselineskip 21pt

\bigskip

\centerline{\Large \bf Large-Scale Outer Rings of Early-type Disk Galaxies}

\bigskip

\centerline{\large I.~P. Kostiuk$^1$ and O.~K. Sil'chenko$^2$}

\noindent
{\it $^1$Special Astrophysical Observatory of the Russian Academy of Sciences, Nizhnij Arkhyz, Russia}

\noindent
{\it $^2$Sternberg Astronomical Institute of the Lomonosov Moscow State University, Moscow, Russia}

\bigskip
\bigskip

\vspace{2mm}
\sloppypar 
\vspace{2mm}

\bigskip

{\small 
\noindent
We have searched for presence of current star
formation in outer stellar rings of early-type disk (S0-Sb) galaxies
by inspecting a representative sample of nearby galaxies with rings 
from the recent Spitzer catalog ARRAKIS (Comer\'on et al. 2014). We have found 
that regular rings (of R-type) reveal young stellar population with the age of less 
than 200~Myr in about half of all the cases, while in the pseudorings (open
rings, R$^{\prime}$), which inhabit only spiral galaxies, current star formation proceeds 
almost always.
}

\clearpage

\section{INTRODUCTION}

Large-scale stellar rings are frequent structural components
of disk galaxies. Vorontsov-Velyaminov~\cite{voron60:Kostyuk_n,voron76:Kostyuk_n}
insisted to consider them as a part of morphology which is just so
important as spiral arms or bars. Indeed, rings of
various scales are present in more than a half of disk
galaxies~\cite{arrakis:Kostyuk_n}. Just like spiral arms, 
these structures may have a smooth regular appearance or be clumpy and
irregular; they may have the galaxy nucleus in the geometrical
center or may be shifted relative to the galactic
center~\cite{fewmadore:Kostyuk_n}. All these features evidently
have an evolutionary sense and are related to the origin of
the ring. Now, the most popular scenaria of the origin
of ring structures in galaxies are the resonance scenario and the impact
one. In the former case, formation of a ring is provoked
by a presence of a bar: the presence of
non-axisymmetric density perturbation (and so the
triaxiality of the gravitational potential), which rotates as a
rigid body demonstrating an angular velocity  constant along the radius 
of the disk, distinguishes dynamically certain radial disk zones
where rotation of the bar proceeds in resonance with the nearly circular 
differential gas rotation. Near these Lindblad resonance radii, the cloud
orbits crowd together, gas gets compressed, triggering the conditions for
intense, very efficient star formation, resulting in the formation
of radial enhancement in the distribution of stars, in other
words,  stellar rings~\cite{ss76:Kostyuk_n,atha82:Kostyuk_n,buta96:Kostyuk_n,schwarz:Kostyuk_n}.
The manifold theory is a modification of the resonance
mechanism, where the flux tubes are produced by stable gas orbits around the
equilibrium points in the bar triaxial potential~\cite{manifold:Kostyuk_n}. 
The impact mechanism~\mbox{\cite{freevauc:Kostyuk_n,theys_spi:Kostyuk_n,fewmadore:Kostyuk_n,atha97:Kostyuk_n}}
involves an external inluence: it is suggested that all very contrast
rings are formed as a result of infall of a companion galaxy from a
highly inclined orbit onto the disk of a galaxy, just near the center.
The impact event results in the disk plane vertical oscillations 
and generates a ring-like density wave running outwards through the galactic disk.    
If the disk contains gas, intense star formation starts at a radius where
the gas is compressed  to the critical density, generating a ring of young stars. 
However, even in the case of a gas-less, purely stellar disk, the impact effect may give 
rise to a transient surface density ring, expanding outwards~\cite{taiwan:Kostyuk_n}. 
It should be noted however that the most
popular resonance mechanism affects only gas. It is because from the dynamical 
point of view, the gas is a collisional system lacking ``elasticity'' of the stellar
components of the disks. Gas clouds cannot traverse the radii where chaotic
orbits dominate -- the resonance areas. There shock fronts develop, and the gas
condenses at a given radius, then igniting star formation.
As a result, the gas ring becomes a stellar ring-like structure of the
disk.

The third possibility to form an outer ring-shaped structure in a
galaxy, which, as will be shown below, we consider to be the most
likely one, is accretion of the outer gas from
a neighboring galaxy through the gravitational tides or
from a cosmological filament during hierarchical assembly of the matter. 
This scenario is not yet very popular among the astronomical community. Once 
it was discussed concerning the discovery of the Hoag object~\cite{schweiz87:Kostyuk_n}, 
as the ring of this galaxy is very massive, and an absolutely round (axisymmetric?)
early-type galaxy is located in its center. Neither the resonance
scenario, nor the impact effect (both affecting only the own gas disk
of the galaxy) would produce such an exotic configuration.
However, as early as in the survey of Buta and Combes~\cite{buta96:Kostyuk_n}, 
it was noted that no other galaxies similar to the Hoag object were found and 
that almost all other outer rings were accompanied by a nonaxisymmetric
distortion of the isophotes in the centers of the galaxies. It has been
concluded that the vast majority of the outer rings, at least those visibly
residing in the main planes of symmetry of the galaxies, have the
resonance origin.

The observational statistics of ring structures in galaxies
was studied extensively. Ronald 
Buta~\mbox{\cite{buta86:Kostyuk_n,butacrocker:Kostyuk_n,buta95:Kostyuk_n}}
made a lot of effort, comparing the metric properties of rings and
bars, and collected arguments in favor of the resonance scenario
for the origin of the most considered rings. However, numerous
cases are known of the presence of rings, sometimes two or
three simultaneously, at different radii in the galaxies without bars, 
and often these galaxies are completely isolated, with no companions or signs
of interaction that unfavored also the impact scenario. 
Sil'chenko and Moiseev~\cite{we2006:Kostyuk_n}
noted that the very presence of the rings in galaxies
without bars and traces of collision with another galaxy indicate
that the origin of rings in galaxies can be very
diverse, including also smooth gas accretion from outside. 
Interestingly, along the Hubble morphological sequence, 
the frequency of bars and that of rings drift oppositely. 
The morphological analysis in the near infrared (at 2--4~$\mu$m) 
has shown that in S0 (also disk!) galaxies, the strong bars can be met
much rarer than even in their nearest neighbors on the Hubble's fork, early spiral
galaxies \mbox{Sa--Sb} ($46\pm 6$\%\ versus
64--93\%~\cite{lauri09:Kostyuk_n}), while the outer ring
structures, in contrast, exist in 60\%\ of S0 galaxies and only in
20\%\ of Sb galaxies~\cite{arrakis:Kostyuk_n}. A large catalog of the
ring-shaped structures in the disks, called
ARRAKIS~\cite{arrakis:Kostyuk_n} was recently compiled
based on the results of a morphological survey of nearby galaxies
with the Spitzer Space Telescope at the wavelengths of 3.6 and
4.5~$\mu$m \cite{buta_s4g:Kostyuk_n}. This catalog describes evidently
only stellar rings, because at the wavelengths of around 4~$\mu$m  we
see the bulk of the old stellar populations. Since all the popular
models relate the appearance of ring-shaped structures to the
gas condensation and subsequent starburst at a certain radius, it would be
interesting to check how often the stellar ring structures in
galaxies reveal the signs of current star formation, especially if
such star formation is not present in the rest of the disk (as
it takes place in the S0 galaxies by definition). The survey of the
morphology of nearby galaxies in the ultraviolet (UV) bands by the 
GALEX space telescope has provided necessary observational data 
to address this issue~\cite{galex:Kostyuk_n}. In the near-UV band, we 
see mostly the stellar population younger than a few hundred million
years old. Thus, the fraction of outer stellar rings visible in the near UV
gives us a rough estimate of the outer ring lifetime or of the time of 
their dissipation. Similar analysis has recently been done by Sebastian
Comer\'on~\cite{comeron:Kostyuk_n} for a sample of inner rings in the
galaxies of the S4G survey. He had picked inner rings from the ARRAKIS catalog 
and identified them on the GALEX maps and in the narrow-band images got around
the emission H$\alpha$ line. His analysis has shown that in the early-type 
disk galaxies (S0--Sab) only 21\% ($\pm 3$\%) of the inner rings do not radiate 
in the UV (though in the far UV, at 1500\,\AA, that reduces the age of the stellar
population searched for in this band even more). Accordingly, the
dissipation time of the inner rings within the frame of hypothesis  of
their resonance origin appears to be at least 200~Myr, that
is comparable to a single rotation period in the central part of a galaxy. 
In the present study we have repeat this analysis for a sample of {\it outer}  
rings in the early-type disk galaxies from the ARRAKIS catalog searching
for recent star formation there.

\section{THE~SAMPLE}

We have selected a sample of early-type disk galaxies, from S0 to Sb,
with the outer ring-shaped structures from the ARRAKIS atlas and catalogue~\cite{arrakis:Kostyuk_n}. 
To classify and describe the galaxies, this atlas used the images from the
Spitzer Survey of Stellar Structure in Galaxies (S4G)~\cite{s4g:Kostyuk_n}, 
which are in public access, and their morphological analysis from \cite{buta_s4g:Kostyuk_n}. 
The S4G sample has the following restrictions: distance of the galaxies
$D<40$~Mpc;  galactic latitude $|b|>30^{\circ}$; the blue
integrated magnitude corrected for dust extinction in our
Galaxy, inclination of the galaxy  disk plane to our line of sight, and the
K-correction $m_{B,{\rm corr}} < 15.5$; angular diameter up to
the 25th isophote in $B$ $>1'$. An outer ring-shaped feature in the ARRAKIS, 
just like other medium-scale features described in this catalog, was identified by inspecting 
the residual images of the galaxies obtained by subtracting a model image consisted of four 
standard components (nucleus, bulge, disk, bar) from the near-infrared galaxy images. 
All the S4G galaxies, more than 2000 nearby ones, were proceeded through this pipeline 
(S4G pipeline 4, P4). From the ARRAKIS we have selected the galaxies with the
outer ring features classified as R, RL, R$_1$, and R$_2$-type rings. 
Further, this group of rings will be considered as a whole and indicated by the letter R; 
consequently, the pseudorings (open rings) will be marked as R$^{\prime}$. Because of
a statistically small number of selected objects, finer separation into
subtypes (R, R$_1$, R$_2$, RL etc) has not been undertaken.  The main
difference between a ring and a pseudoring is a shape of
the former as a closed outer feature with the surface brightness dip
between the inner part of the galaxy and the ring. 
If in the ARRAKIS a galaxy was listed as possessing two outer ring-shaped structures,  
we took into account only the outermost of the rings. Just like in the ARRAKIS, the
galaxies with the outer isophote ellipticity exceeding 0.5 (edge-on disks) 
were not included into the sample because of large uncertainty in the classification. 
The S4G full sample includes 2331 galaxies of all morphological types~\cite{s4g:Kostyuk_n}, 
and according to the ARRAKIS, 277 of them have an outer ring-shaped feature R or R$^{\prime}$ 
(for 18 of them two outer ring-shaped features are noted). After the additional selection 
by the morphological type (202 galaxies of 277 have the S0--Sb morphological types) 
and by the ellipticity ($1-b/a$ less than 0.5), 145 galaxies were left in our sample.

Let us note that many years ago Kostyuk~\cite{kostuk75:Kostyuk_n} compiled a list of 
143~ring-shaped galaxies selected `by eye' from the photographic copies of the Palomar 
Observatory Sky Survey (POSS) maps of all the northern sky. If we apply additional
restrictions on the size, larger than 1.0$^{\prime}$, and on the galactic latitude, 
$|b|>30^{\circ}$, in that list we are left with 51~galaxies.  And only 18 galaxies from
the list of~\cite{kostuk75:Kostyuk_n} have Hubble velocities lower
than 3000~km\,s$^{-1}$; 11 of them are included into the ARRAKIS. We admit that 
the visual inspection is less effective than the pipeline reduction, however,  
the above mentioned list includes some very interesting galaxies not
included into the ARRAKIS, a detailed study of which is planned by us in the future.

\begin{figure}[tbp!!!]
\vspace{4.6mm}
\includegraphics[width=0.95\columnwidth]{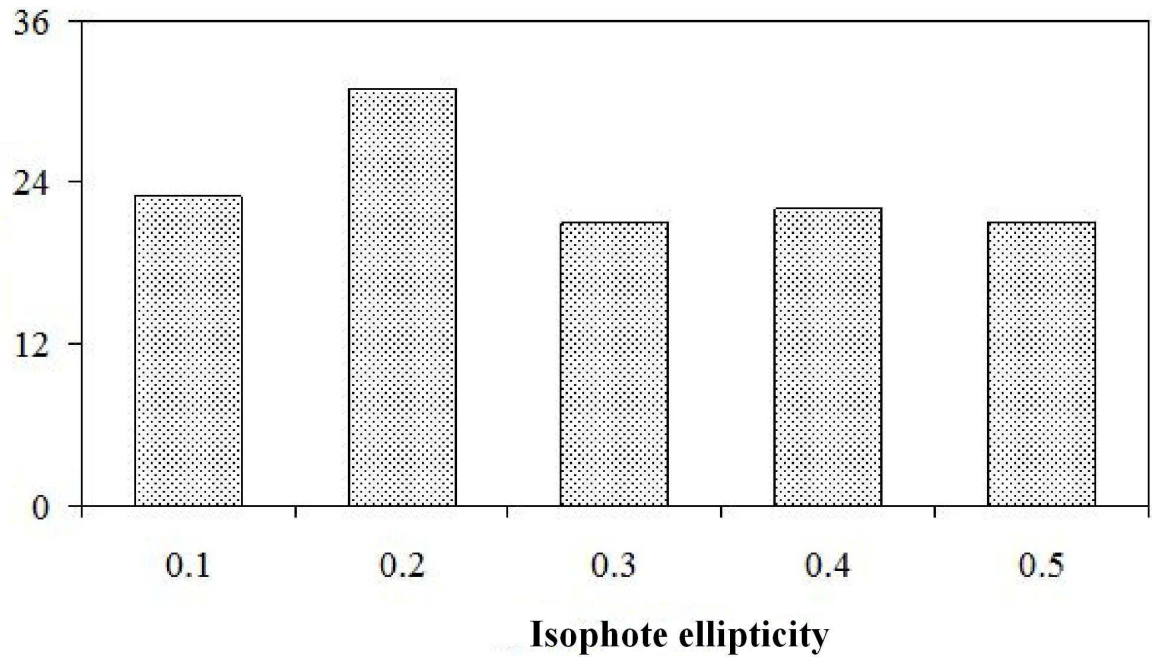}
\vspace{-4.5mm}
\caption{Distribution of the selected ring-shaped galaxies by the
outer isophote ellipticity. }\label{fig1:Kostyuk_n}
\end{figure}

\section{CURRENT STAR FORMATION IN~THE~OUTER RINGS}

All the 145 galaxies of our sample have been retrieved in the data
archive of the GALEX space telescope  \footnote{\tt
http://galex.stsci.edu/GR6/}. Rather massive, and so young,
stars having lifetimes of up to 200~Myr dominate in the near-UV
spectral range~\cite{martin05:Kostyuk_n,ken_evans:Kostyuk_n}. So to study the
presence of young stars, we searched for the images in the near-UV band (NUV) at \mbox{1770--2730}~\AA\ 
expressed as intensity maps. The results of our search have revealed that the data for
nine galaxies from the list are missing  in the GALEX survey, for
16~galaxies their GALEX images clearly show a spiral structure outside the outer ring,
and bright stars projected near to two galaxies prevent the analysis of their faint outer
UV-structures. Hence, these 18 galaxies were excluded from our analysis.

Finally, the list for studying the UV morphologies of the outer
rings in the early-type disk galaxies consists of 118 galaxies.
Figure~1 shows the distribution of these ring-shaped galaxies by
the ellipticity of the outer isophotes according to the S4G survey
data.
This distribution is fairly flat, and we assure that there is
no bias through the inclination of the galaxy, only after excluding
the disks inclined at an angle of more than $60^{\circ}$ ($b/a>0.5$). The
fractions of various morphological types among the ring-possessing and
pseudoring-possessing galaxies differ (Fig.~2). Among the galaxies with
pseudorings, R$^{\prime}$, there are no S0 galaxies. The galaxies with the
closed R-type rings, on the contrary, are dominated by the S0 type, and 
the presence of the morphological type Sab--Sb is several times smaller 
than that in the galaxies with pseudorings R$^{\prime}$.

\begin{figure}[tbp!!!]
 \vspace{-1mm}
\includegraphics[width=0.95\columnwidth]{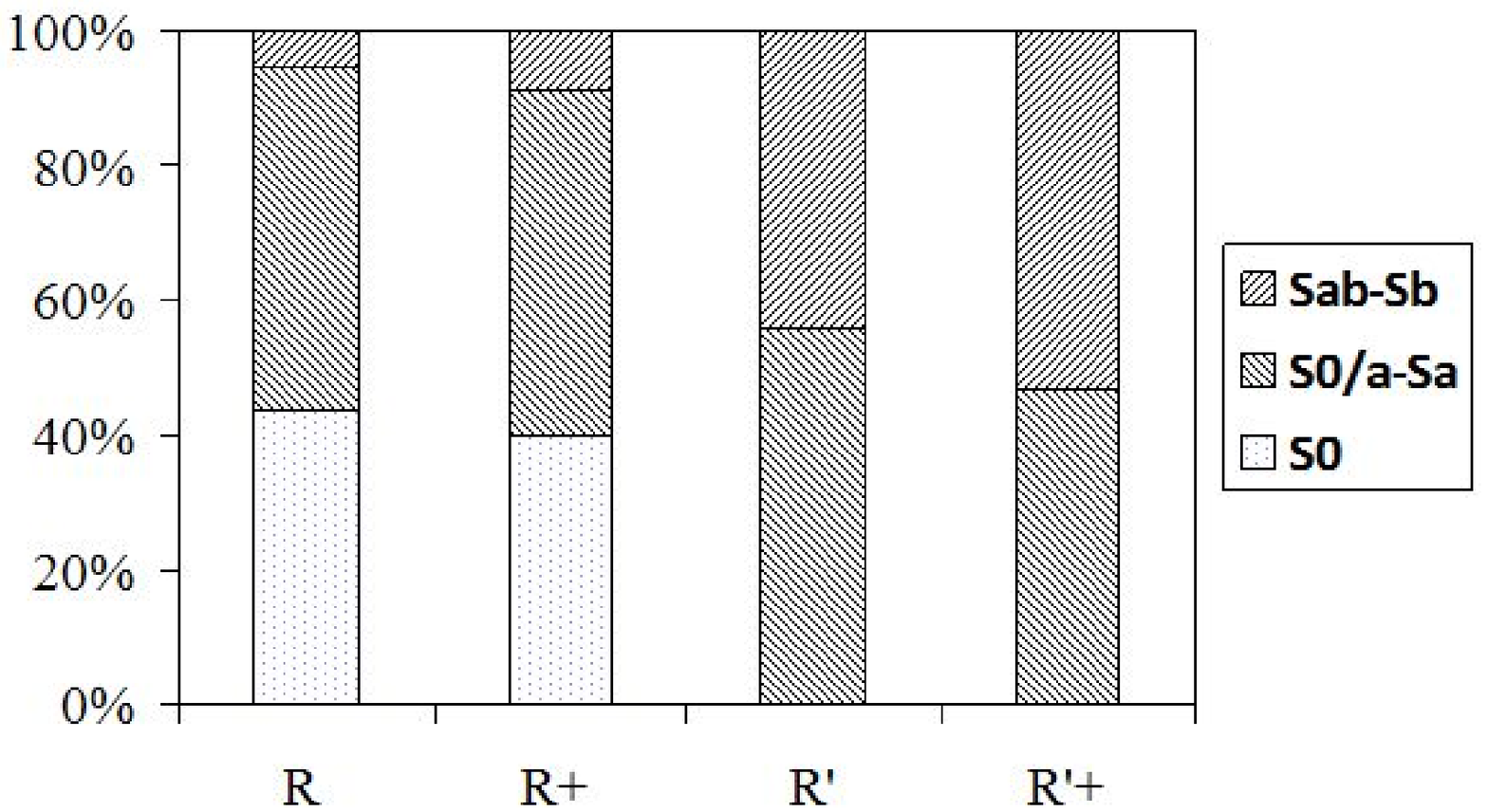}
 \vspace{1.5mm}
\caption{Fraction of galaxies of different morphological types
(S0, S0/a--Sa, Sab--Sb) among the ring-shaped galaxies. By the big `R' the
galaxies with closed rings are marked, R$+$ are the ring galaxies with 
the NUV radiation in the ring, R$^{\prime}$ are the pseudoring galaxies,
\mbox{R$^{\prime}+$ are the pseudoring} galaxies with UV radiation
in the ring.}\label{fig2:Kostyuk_n}
\end{figure}

We have estimated quantitatively the significance of the NUV
flux in the outer rings of the galaxies by calculating the signal-to-noise ratio
in the fixed area of the images accumulated by the GALEX telescope. An ellipse with
the major and minor axes $D_r$ and $d_r$ (according to \mbox{ARRAKIS})
was superposed onto the NUV image of every galaxy. 

\renewcommand{\baselinestretch}{1.05}
\begin{longtable}{l|l|c|c|l|c|c|c}
\caption{A list of early-type disk galaxies with the outer ring-shaped  structures}\\
\hline
\multicolumn{1}{c|}{Galaxy} &  \multicolumn{1}{c|}{Type}  & Ring size  & Disk size & \multicolumn{1}{c|}{Ring type}  & Is UV present? & $k$ & Notes \\[-5pt]
 &  & (\mbox{ARRAKIS}),& (RC3),& (\mbox{ARRAKIS})& & & \\
 & & arcmin & arcmin  &&&&\\
 \hline
\multicolumn{1}{c|}{(1)} &\multicolumn{1}{c|}{(2)} & (3) & (4) & \multicolumn{1}{c|}{(5)} & (6) & (7) & (8) \\
\hline\endfirsthead
\caption{(Contd.)}\\
\hline
\multicolumn{1}{c|}{Galaxy} &  \multicolumn{1}{c|}{Type}  & Ring size  & Disk size & \multicolumn{1}{c|}{Ring type}  & Is UV present? & $k$ & Notes \\[-5pt]
 &  & (\mbox{ARRAKIS}),& (RC3),& (\mbox{ARRAKIS})& & & \\
 & & arcmin & arcmin  &&&&\\
 \hline
\multicolumn{1}{c|}{(1)} &\multicolumn{1}{c|}{(2)} & (3) & (4) & \multicolumn{1}{c|}{(5)} & (6) & (7) & (8) \\
\hline\endhead\hline
\endfoot
\endlastfoot
NGC\,210   & SABab  & $4.08\times 2.90$  & $5.01\times 3.31$ & ~~~~~~~~R$_2^{\prime}$L & $+$ & 8 & 2 \\
NGC\,254   & SAB0   & $1.59\times 1.25$  & $2.46\times 1.52$ & ~~~~~~~~R & --    & -- & -- \\
NGC\,474   & SAB0/a & $2.23\times 2.07$  & $7.08\times 6.30$ & ~~~~~~~~R$^{\prime}$ & -- & -- & -- \\
NGC\,615   & SABa   & $2.26\times 0.70$  & $3.63\times 1.45$ & ~~~~~~~~R$_2^{\prime}$ & $+$ &  4 & 2 \\
NGC\,691   & SAab   & $2.74\times 1.93$  & $3.47\times 2.63$ & ~~~~~~~~R & $+$ &   3& 2 \\
NGC\,718   & SABa   & $1.42\times 1.09$  & $2.34\times 2.04$ & ~~~~~~~~R$^{\prime}$ &  -- &    -- & -- \\
NGC\,986   & SBab   & $3.52\times 2.59$  & $3.89\times 2.96$ & ~~~~~~~~R$^{\prime}$ &   $+$ & 10 & 1 \\
NGC\,1022  & SAB0/a & $2.08\times 1.66$  & $2.40\times 1.99$ & ~~~~~~~~RL &   -- &    -- &    -- \\
NGC\,1068  & SAa    & $5.81\times 4.89$  & $7.08\times 6.02$ & ~~~~~~~~R & $+$ & 2 & 2 \\
NGC\,1258  & SABa:  & $0.78\times 0.35$  & $1.35\times 0.93$ & ~~~~~~~~R$^{\prime}$ &  $+$ &   20 & 2\\
NGC\,1291  & SAB0   & $8.47\times 7.08$  & $9.77\times 8.11$ & ~~~~~~~~R & $+$ &   5 & 2 \\
NGC\,1300  & SBb    & $5.63\times 4.90$  & $6.17\times 4.07$ & ~~~~~~~~R$^{\prime}$ &  $+$ &   8 & 2 \\
NGC\,1326  & SAB0   & $2.83\times 1.87$  & $3.89\times 2.88$ & ~~~~~~~~R$_1$ & $+$ &   5 & 2 \\
ESO\,548-23& SA0    & $0.66\times 0.35$  & $1.05\times 0.47$ & ~~~~~~~~RL &    $+$ &   3 & 3 \\
NGC\,1350  & SAB0/a & $5.35\times 2.60$  & $5.25\times 2.83$ & ~~~~~~~~R &    $+$ &   7 & 2 \\
NGC\,1357  & SA0/a  & $2.74\times 2.20$  & $2.81\times 1.95$ & ~~~~~~~~R$^{\prime}$L & -- & -- & -- \\
NGC\,1398  & SBa    & $4.64\times 3.17$  & $7.08\times 5.38$ & ~~~~~~~~R & $+$ &   9 & 2 \\
NGC\,1433  & SBa    & $6.26\times 4.48$  & $6.46\times 5.88$ & ~~~~~~~~R$_1^{\prime}$ & $+$ &  3 & 1, 2 \\
NGC\,1436  & SABab  & $1.59\times 0.99$  & $2.95\times 2.01$ & ~~~~~~~~R$^{\prime}$ & $+$ &    16 &    2 \\
NGC\,1452  & SB0/a  & $2.54\times 1.43$  & $2.24\times 1.48$ & ~~~~~~~~RL & -- & -- & -- \\
IC\,1993   & SABab  & $1.47\times 1.44$  & $2.46\times 2.14$ & ~~~~~~~~R$^{\prime}$ &   $+$ &   14 & 2 \\
NGC\,1533  & SB0    & $1.66\times 1.43$  & $2.76\times 2.34$ & ~~~~~~~~RL &    $+$ &   3 & 3 \\
NGC\,1566  & SABb   & $7.55\times 6.34$  & $8.32\times 6.57$ & ~~~~~~~~R$_1^{\prime}$ &    $+$ &   6 & 2 \\
NGC\,1640  & SBa    & $1.74\times 1.55$  & $2.63\times 2.05$ & ~~~~~~~~R$^{\prime}$ & $+$ & 6 & 2 \\
NGC\,1808  & SABa   & $6.38\times 4.62$  & $6.46\times 3.87$ & ~~~~~~~~R$_1$ & $+$ &   3 & 1, 2 \\
NGC\,2681  & SAB0/a & $2.35\times 2.13$  & $3.63\times 3.30$ & ~~~~~~~~R &    $+$ &   3 & 3 \\
NGC\,2685  & S0     & $4.08\times 1.80$  & $4.47\times 2.37$ & ~~~~~~~~R & -- &    -- & -- \\
NGC\,2712  & SABab  & $2.73\times 1.16$  & $2.88\times 1.59$ & ~~~~~~~~R$^{\prime}$ &    $+$ &   3 & 2 \\
NGC\,2780  & SBa    & $0.67\times 0.49$  & $0.89\times 0.66$ & ~~~~~~~~R$^{\prime}$ &  $+$ &   8 & 3 \\
NGC\,2859  & SAB0   & $3.42\times 2.73$  & $4.26\times 3.80$ & ~~~~~~~~R &     $+$ &   2 & -- \\
NGC\,2893  & SAB0   & $0.87\times 0.69$  & $1.12\times 1.02$ & ~~~~~~~~RL &    $+$ &   5 & 1, 2 \\
NGC\,2962  & SAB0   & $2.13\times 1.42$  & $2.63\times 1.95$ & ~~~~~~~~R & $+$ &   4 &  2 \\
NGC\,3166  & SB0    & $4.11\times 1.73$  & $4.79\times 2.35$ & ~~~~~~~~RL & $+$ &  2 & 3 \\
NGC\,3185  & SABa   & $2.51\times 1.65$  & $2.35\times 1.59$ & ~~~~~~~~RL &    -- &    -- & -- \\
NGC\,3248  & SA0    & $1.77\times 1.10$  & $2.51\times 1.56$ & ~~~~~~~~RL &    -- &    -- & -- \\
NGC\,3266  & SB0    & $0.79\times 0.60$  & $1.55\times 1.32$ & ~~~~~~~~RL & -- & -- & -- \\
NGC\,3368  & SAB0   & $5.94\times 3.57$  & $7.58\times 5.23$ & ~~~~~~~~RL &    $+$ &   4 & 2 \\
NGC\,3380  & SAB0/a & $1.35\times 1.29$  & $1.70\times 1.34$ & ~~~~~~~~RL & $+$ & 2 & 1 \\
NGC\,3489  & SB0    & $1.53\times 0.45$  & $3.55\times 2.02$ & ~~~~~~~~R & $+$ &   8 & 2 \\
NGC\,3504  & SABa   & $2.03\times 1.89$  & $2.69\times 2.10$ & ~~~~~~~~R$_1^{\prime}$ &    $+$ & 6 & 2 \\
NGC\,3507  & SABb   & $2.34\times 2.11$  & $3.39\times 2.88$ & ~~~~~~~~R$^{\prime}$ & $+$ & 8 & 2 \\
NGC\,3583  & SABab  & $1.96\times 1.04$  & $2.81\times 1.83$ & ~~~~~~~~R$^{\prime}$ &  $+$ &   13 &    2 \\
NGC\,3637  & SB0    & $1.44\times 1.12$  & $1.59\times 1.55$ & ~~~~~~~~RL &    -- &    -- & -- \\
NGC\,3675  & SAb    & $2.42\times 0.89$  & $5.89\times 3.12$ & ~~~~~~~~R$^{\prime}$ &  $+$ &   10 & 2 \\
IC\,2764   & SA0    & $0.84\times 0.70$  & $1.62\times 1.41$ & ~~~~~~~~R & $+$ & 6 & 2 \\
NGC\,3687  & SABab  & $1.43\times 1.34$  & $1.91\times 1.91$ & ~~~~~~~~RL &    $+$ &   6 & 2 \\
NGC\,3786  & SA0/a  & $1.77\times 0.79$  & $2.19\times 1.29$ & ~~~~~~~~R & -- &    -- & -- \\
NGC\,3892  & SAB0   & $2.55\times 2.20$  & $2.95\times 2.24$ & ~~~~~~~~RL &    -- &    -- & -- \\
NGC\,3941  & SB0    & $1.78\times 0.94$  & $3.47\times 2.29$ & ~~~~~~~~R & $+$ &   3 & 3 \\
NGC\,4045  & SABab  & $1.91\times 0.95$  & $2.69\times 1.86$ & ~~~~~~~~R$_1^{\prime}$L & $+$ & 4 & 2 \\
NGC\,4050  & SABa   & $3.36\times 2.01$  & $3.09\times 2.10$ & ~~~~~~~~RL &    -- &    -- & -- \\
NGC\,4102  & SABab  & $1.24\times 0.67$  & $3.02\times 1.72$ & ~~~~~~~~R$^{\prime}$ &  $+$ &   20 & 2 \\
IC\,3102   & SAB0/a & $2.76\times 1.45$  & $2.57\times 1.36$ & ~~~~~~~~R$^{\prime}$L & -- & -- &    -- \\
NGC\,4245  & SB0    & $2.65\times 1.89$  & $2.88\times 2.19$ & ~~~~~~~~RL &    -- &    -- & -- \\
NGC\,4286  & SA0    & $0.78\times 0.55$  & $1.59\times 1.00$ & ~~~~~~~~RL &    $+$ &   4 & 3 \\
NGC\,4314  & SBa    & $3.72\times 3.03$  & $4.17\times 3.71$ & ~~~~~~~~R$_1^{\prime}$ &    -- &    -- & -- \\
NGC\,4355  & SAB0/a & $0.97\times 0.49$  & $1.45\times 0.71$ & ~~~~~~~~R$^{\prime}$L & $+$ & 2 & 3 \\
NGC\,4369  & SB0/a  & $1.42\times 1.30$  & $2.09\times 2.05$ & ~~~~~~~~R & $+$ &   2 & 3 \\
NGC\,4378  & SAa    & $2.93\times 2.45$  & $2.88\times 2.68$ & ~~~~~~~~R$^{\prime}$ &  $+$ &   3 & 2, 3 \\
NGC\,4380  & SAab   & $2.02\times 1.07$  & $3.47\times 1.91$ & ~~~~~~~~R & $+$ &   7 & 2, 3 \\
NGC\,4394  & SB0/a  & $2.74\times 2.45$  & $3.63\times 3.23$ & ~~~~~~~~R & $+$ &   9 & 2 \\
NGC\,4405  & SABa   & $1.05\times 0.66$  & $1.78\times 1.16$ & ~~~~~~~~R & $+$ &   3 & 3 \\
NGC\,4407  & SBab   & $2.81\times 1.52$  & $2.35\times 1.52$ & ~~~~~~~~R$^{\prime}$L & -- & -- & -- \\
NGC\,4424  & SB0/a  & $3.18\times 1.53$  & $3.63\times 1.82$ & ~~~~~~~~R$_2 ^{\prime}$L & -- &   -- &    -- \\
NGC\,4450  & SABa   & $3.43\times 2.20$  & $5.25\times 3.88$ & ~~~~~~~~R$^{\prime}$ &  $+$ & 3 & 2 \\
NGC\,4454  & SAB0/a & $2.00\times 1.83$  & $2.00\times 1.70$ & ~~~~~~~~RL &   -- &    -- & -- \\
NGC\,4457  & SAB0   & $4.04\times 3.76$  & $2.69\times 2.29$ & ~~~~~~~~R & -- &    -- & -- \\
NGC\,4579  & SBa    & $4.33\times 3.10$  & $5.89\times 4.65$ & ~~~~~~~~RL & -- & -- & -- \\
NGC\,4580  & SAa    & $1.86\times 1.26$  & $2.09\times 1.63$ & ~~~~~~~~R$^{\prime}$ & -- & -- & -- \\
NGC\,4593  & SBa    & $3.54\times 2.51$  & $3.89\times 2.88$ & ~~~~~~~~R$^{\prime}$ & $+$ & 4 & 2 \\
NGC\,4596  & SB0/a  & $3.40\times 2.66$  & $3.98\times 2.95$ & ~~~~~~~~RL &    -- &    -- & -- \\
NGC\,4639  & SBab   & $2.36\times 1.38$  & $2.76\times 1.87$ & ~~~~~~~~R$^{\prime}$ &  $+$ & 7 & 2 \\
NGC\,4659  & SAB0   & $1.04\times 0.64$  & $1.74\times 1.25$ & ~~~~~~~~R & $+$ & 2 & 3 \\
NGC\,4691  & S0/a   & $2.79\times 2.15$  & $2.81\times 2.28$ & ~~~~~~~~R$^{\prime}$L & -- &    -- & -- \\
NGC\,4698  & SA0/a  & $7.91\times 2.73$  & $3.98\times 2.47$ & ~~~~~~~~R & $+$ & 2 & -- \\
NGC\,4699  & SB0/a  & $1.95\times 1.48$  & $3.80\times 2.62$ & ~~~~~~~~R$^{\prime}$ &  $+$ & 7 & 2 \\
NGC\,4736  & SABa   & $10.58\times 8.68$ & $11.22\times 9.09$& ~~~~~~~~R & $+$ & 2 & -- \\
NGC\,4750  & SAa    & $1.52\times 1.33$  & $2.04\times 1.86$ & ~~~~~~~~R$^{\prime}$ & $+$ & 18 & 2 \\
NGC\,4772  & SA0/a  & $3.85\times 1.93$  & $3.38\times 1.69$ & ~~~~~~~~R$^{\prime}$ &  -- &    -- & -- \\
NGC\,4795  & SBa    & $1.38\times 1.13$  & $1.86\times 1.58$ & ~~~~~~~~R$^{\prime}$ & -- & -- & -- \\
NGC\,4800  & SAa    & $1.22\times 0.97$  & $1.58\times 1.17$ & ~~~~~~~~R$^{\prime}$ & $+$ & 8 & 3 \\
NGC\,4826  & SAa    & $7.17\times 3.10$  & $10.00\times 5.40$& ~~~~~~~~R$^{\prime}$ &  $+$ &   2 & 3 \\
NGC\,4856  & SB0    & $2.44\times 0.65$  & $4.26\times 1.19$ & ~~~~~~~~RL & -- & -- & -- \\
NGC\,4880  & SAB0   & $2.05\times 1.48$  & $3.16\times 2.47$ & ~~~~~~~~RL &    -- &    -- & -- \\
NGC\,4941  & SA0/a  & $3.33\times 2.36$  & $3.63\times 1.96$ & ~~~~~~~~RL &    $+$ &   2 & 1, 2 \\
NGC\,4984  & SAB0/a & $5.07\times 2.82$  & $2.76\times 2.18$ & ~~~~~~~~R$^{\prime}$ & -- &    -- & -- \\
IC\,863    & SBb    & $0.51\times 0.30$  & $1.82\times 1.20$ & ~~~~~~~~R$^{\prime}$ &  $+$ &   4 & 3 \\
IC\,4214   & SAB0/a & $2.03\times 1.25$  & $2.24\times 1.28$ & ~~~~~~~~R$_1$ &    $+$ &   3 & 2 \\
NGC\,5101  & SB0/a  & $5.31\times 4.61$  & $5.37\times 4.56$ & ~~~~~~~~R$_2 ^{\prime}$ & $+$ & 2 & 2 \\
NGC\,5134  & SAB0/a & $3.53\times 2.99$  & $2.76\times 1.65$ & ~~~~~~~~R &    -- & -- & -- \\
NGC\,5375  & SBa    & $2.24\times 1.82$  & $3.24\times 2.75$ & ~~~~~~~~R$^{\prime}$ & $+$ &    4 & 2 \\
NGC\,5377  & SAB0/a & $4.05\times 2.03$  & $3.72\times 2.08$ & ~~~~~~~~R$_1$ &    $+$ &   2 & 2 \\
NGC\,5534  & SBa    & $1.21\times 0.83$  & $1.41\times 0.83$ & ~~~~~~~~R$^{\prime}$L & $+$ & 6 & 2 \\
NGC\,5602  & SA0    & $1.05\times 0.54$  & $1.45\times 0.77$ & ~~~~~~~~RL & -- & -- & -- \\
NGC\,5678  & SAb    & $2.31\times 1.17$  & $3.31\times 1.62$ & ~~~~~~~~R$^{\prime}$ & $+$ & 6 & 2 \\
NGC\,5701  & SB0/a  & $3.21\times 2.72$  & $4.27\times 4.05$ & ~~~~~~~~R$_1 ^{\prime}$ &   $+$ &   9 & 2 \\
NGC\,5713  & SBab:  & $1.67\times 1.48$  & $2.76\times 2.45$ & ~~~~~~~~R$^{\prime}$ &  $+$ &   2 & 3 \\
NGC\,5728  & SB0/a  & $3.61\times 2.34$  & $3.09\times 1.76$ & ~~~~~~~~R$_1$ & -- & -- & -- \\
NGC\,5750  & SAB0/a & $2.70\times 1.30$  & $3.02\times 1.60$ & ~~~~~~~~RL &   -- & -- & -- \\
NGC\,5757  & SBab   & $1.35\times 1.19$  & $2.00\times 1.62$ & ~~~~~~~~R$_2 ^{\prime}$ &    $+$ & 3 & 2 \\
NGC\,5806  & SABab  & $2.68\times 1.46$  & $3.09\times 1.58$ & ~~~~~~~~R$^{\prime}$ &  $+$ &   2 & 2 \\
NGC\,5850  & SBab   & $4.00\times 3.31$  & $4.27\times 3.71$ & ~~~~~~~~R$^{\prime}$ &  $+$ &   2 & 2 \\
NGC\,5957  & SBa    & $2.38\times 2.08$  & $2.82\times 2.62$ & ~~~~~~~~R$^{\prime}$ & $+$ & 3 & 2 \\
NGC\,6012  & SBab   & $2.57\times 2.32$  & $2.09\times 1.50$ & ~~~~~~~~R$^{\prime}$ & $+$ & 4 & 2 \\
NGC\,6217  & SBb    & $2.72\times 2.46$  & $3.02\times 2.51$ & ~~~~~~~~R$^{\prime}$  & $+$ &   3 & 1 \\
NGC\,6340  & SA0/a  & $1.80\times 1.52$  & $3.24\times 2.94$ & ~~~~~~~~R & $+$ &   2 & -- \\
NGC\,7051  & SABb   & $1.12\times 0.97$  & $1.32\times 1.09$ & ~~~~~~~~R$_2 ^{\prime}$ &   $+$ &   2 & 1 \\
NGC\,7098  & SAB0/a & $3.63\times 2.26$  & $4.07\times 2.65$ & ~~~~~~~~R &    $+$ &   3 & 2 \\
NGC\,7140  & SABab  & $3.72\times 2.60$  & $4.17\times 3.00$ & ~~~~~~~~R$^{\prime}$ &  $+$ &   2 & 2 \\
NGC\,7191  & SABb   & $0.83\times 0.37$  & $1.59\times 0.55$ & ~~~~~~~~R$^{\prime}$ &  $+$ &   8 & 2 \\
NGC\,7219  & SABa   & $1.36\times 0.83$  & $1.74\times 1.04$ & ~~~~~~~~R$_2 ^{\prime}$ &   $+$ &   10 & 2 \\
IC\,1438   & SAB0/a & $1.77\times 1.32$  & $2.40\times 2.04$ & ~~~~~~~~R$_1$ &    $+$ &   3 & 2 \\
NGC\,7421  & SBab   & $1.68\times 1.54$  & $2.04\times 1.82$ & ~~~~~~~~R$^{\prime}$ &  $+$ &   6 & 2 \\
IC\,5267   & SA0/a  & $4.65\times 3.31$  & $5.25\times 3.88$ & ~~~~~~~~RL & $+$ & 3 & 2 \\
NGC\,7479  & SBb    & $2.77\times 2.13$  & $4.07\times 3.10$ & ~~~~~~~~R$^{\prime}$ & $+$ & 20 & 2 \\
NGC\,7552  & SBa    & $3.08\times 2.52$  & $3.39\times 2.68$ & ~~~~~~~~R$_1 ^{\prime}$ & $+$ & 10 & 2 \\
NGC\,7724  & SABa   & $0.88\times 0.54$  & $1.45\times 1.00$ & ~~~~~~~~R$^{\prime}$ & $+$ & 5 & 2 \\
NGC\,7731  & SABa   & $1.09\times 0.85$  & $1.41\times 1.12$ & ~~~~~~~~R$_2 ^{\prime}$ &   $+$ &   2 & 2 \\
\hline
\end{longtable}
\renewcommand{\baselinestretch}{1.0}

\begin{figure*}
 \vspace{2mm}
\begin{tabular}{c c c}
 \includegraphics[height=0.30\columnwidth]{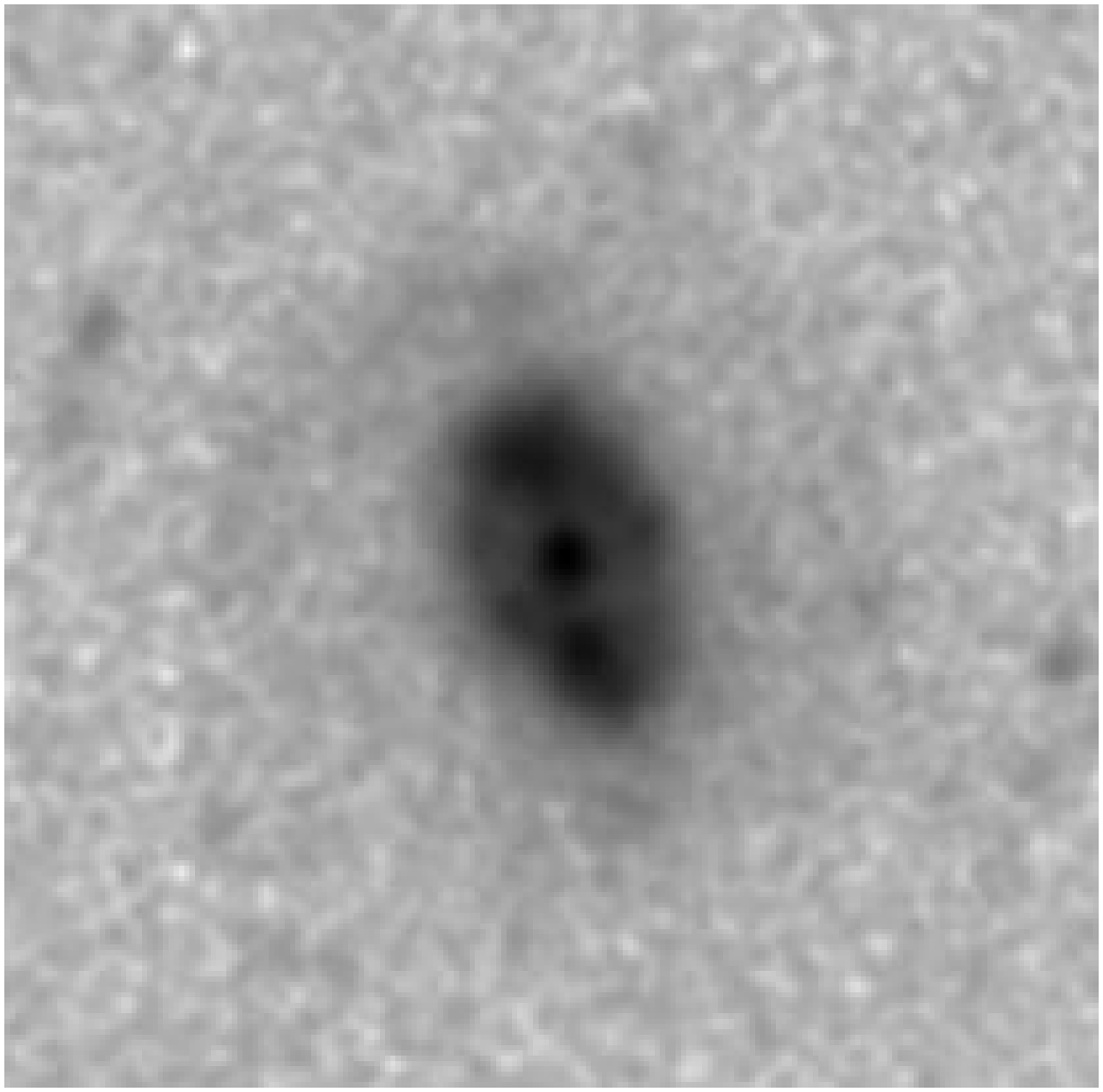} &
 \includegraphics[height=0.30\columnwidth]{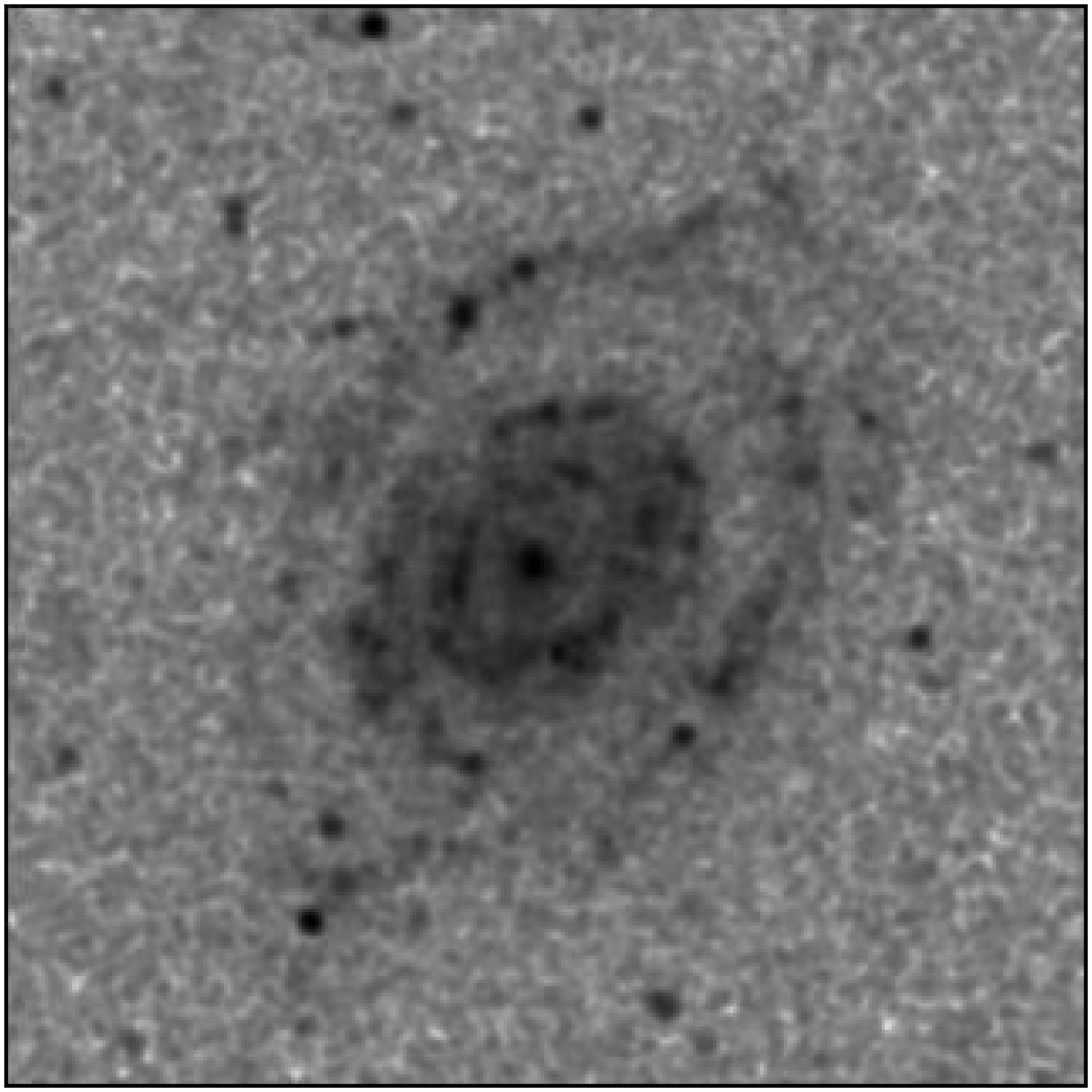} &
 \includegraphics[height=0.30\columnwidth]{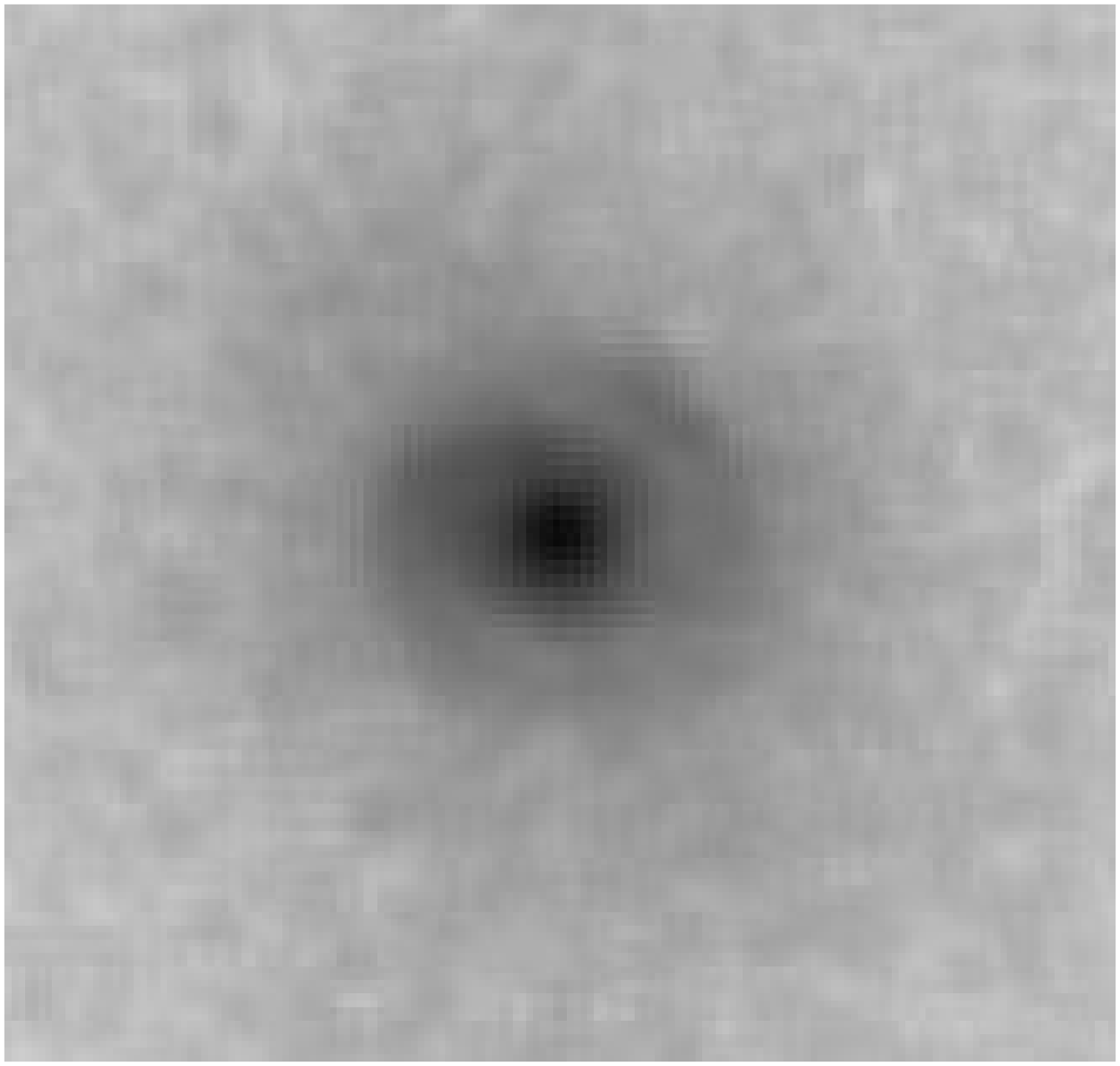} \\
\end{tabular}
\caption{Examples of the GALEX NUV-images for the
galaxies with the ring structures (type R), which correspond to the different
ring subtypes given in the last column of the Table~1. The left plot: the ring
subtype~1, unclosed (NGC\,3380, the field-of-view size is 2.5$^{\prime}$); the
central plot: the ring subtype~2, clumpy (IC\,5267, the field-of-view size is
$8^{\prime}$); the right plot: the ring subtype~3, a filled disk
(NGC\,2681, the field-of-view size is $2^{\prime}$). }\label{fig3:Kostyuk_n}
\end{figure*}

\noindent The outer rings are generally faint features of galaxies.
The galaxies in which an external, with respect to the ring,
extension of the structureless disk is visible in
the NUV band were not excluded from our sample. To estimate a
brightness of the ring in the NUV band, the GALEX image of every
galaxy was smoothed by a 5--10-pixel window. If the
mean count value per pixel in the ring
area exceeded two values of the surrounding sky
background, then the galaxy was marked as having NUV radiation in
its outer ring-shaped feature.

Our estimates of the sky background in the GALEX intensity maps over
vicinity of 117 galaxies of the list above give an average value of
\mbox{$0.00274$~cts\,s$^{-1}$} per pixel, which corresponds to
the average sky surface brightness of $27.36^{\rm m}/\Box''$ in the NUV
band in the~AB system.
The sky background around galaxies varies in the range
of  \mbox{$0.0020$--$0.0042$~cts\,s$^{-1}$}, with the standard
deviation of $0.00049$~cts\,s$^{-1}$. One galaxy (NGC\,1068) was
prominent as concerning its sky NUV background and was not included
into the background estimation sample, since only for
this galaxy the value of the sky background is
$0.0070$~cts\,s$^{-1}$. For 36 galaxies from the list with faint
(maximum three values of sky background) ring-shaped features the
average background is~~$0.00278$~cts\,s$^{-1}$  per pixel falling
into the same range of values and standard deviations, as indicated
above for the whole sample. This corresponds to the sky surface
brightness in the NUV band of $27.35^{\rm m}/\Box''$.

The list consisting of 118 galaxies with the outer ring-shaped structures 
which we have studied is presented in Table~1. The table contains in its
columns as follows: the name of the galaxy; the galaxy
classification taken by us from \mbox{ARRAKIS} without details; 
major and minor axes of the outer ring-shaped structure  $D_r$ and $d_r$ in
arcminutes (from \mbox{ARRAKIS}); the major  2$a$ and minor 2$b$
axes of isophotes of the galactic disk at the surface brightness
level of $25^{\rm m}/\Box''$ in the blue $B$-band (from the RC3
catalog~\cite{rc3:Kostyuk_n}); the  type of the ring-shaped
structure from \mbox{ARRAKIS}; the presence or absence of a
ring-shaped UV signal in the NUV band according to
the GALEX (marked with ``$+$'' or \mbox{``$-$''} respectively);
the $k$ coefficient -- an approximate average value of the UV flux 
in the ring per pixel in the units of the sky background according to the
GALEX (if $k<2$, the minus is in this column); the notes to
the shape of the ring structure as seen in the NUV:
\mbox{1---incomplete,} \mbox{2---clumpy,} \mbox{3---a} filled
disk. Thus, a galaxy without the notes and with the ``$+$''--sign in
the column~6 has indeed a rather uniform ring-shaped structure in the UV,
visible at all azimuths. Figures~3~and~4 show the examples
of all the kinds of ring-shaped structures described in the notes
of Table~1.

\begin{figure}
 \vspace{2mm}
\begin{tabular}{c c}
 \includegraphics[width=0.45\columnwidth]{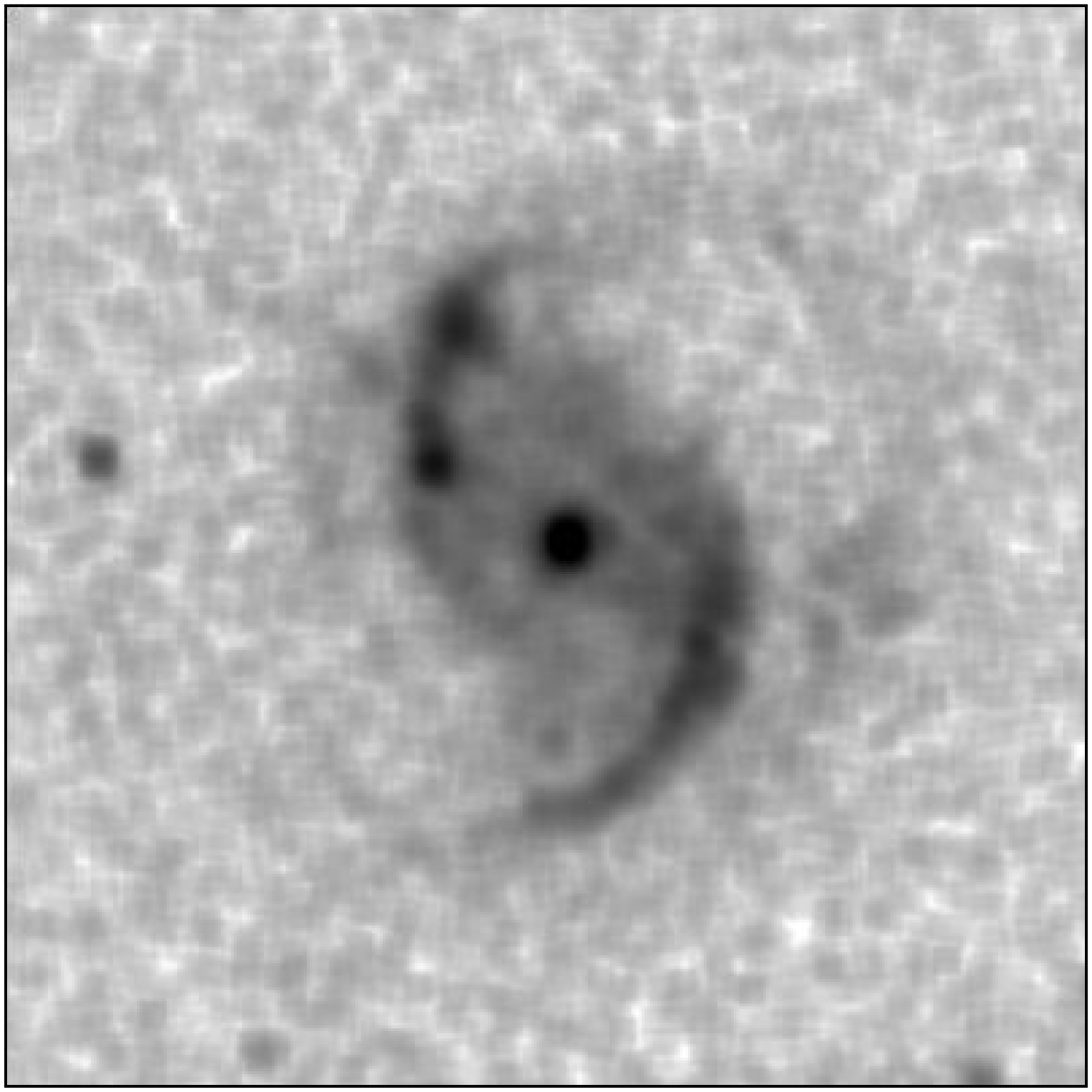} &
 \includegraphics[width=0.45\columnwidth]{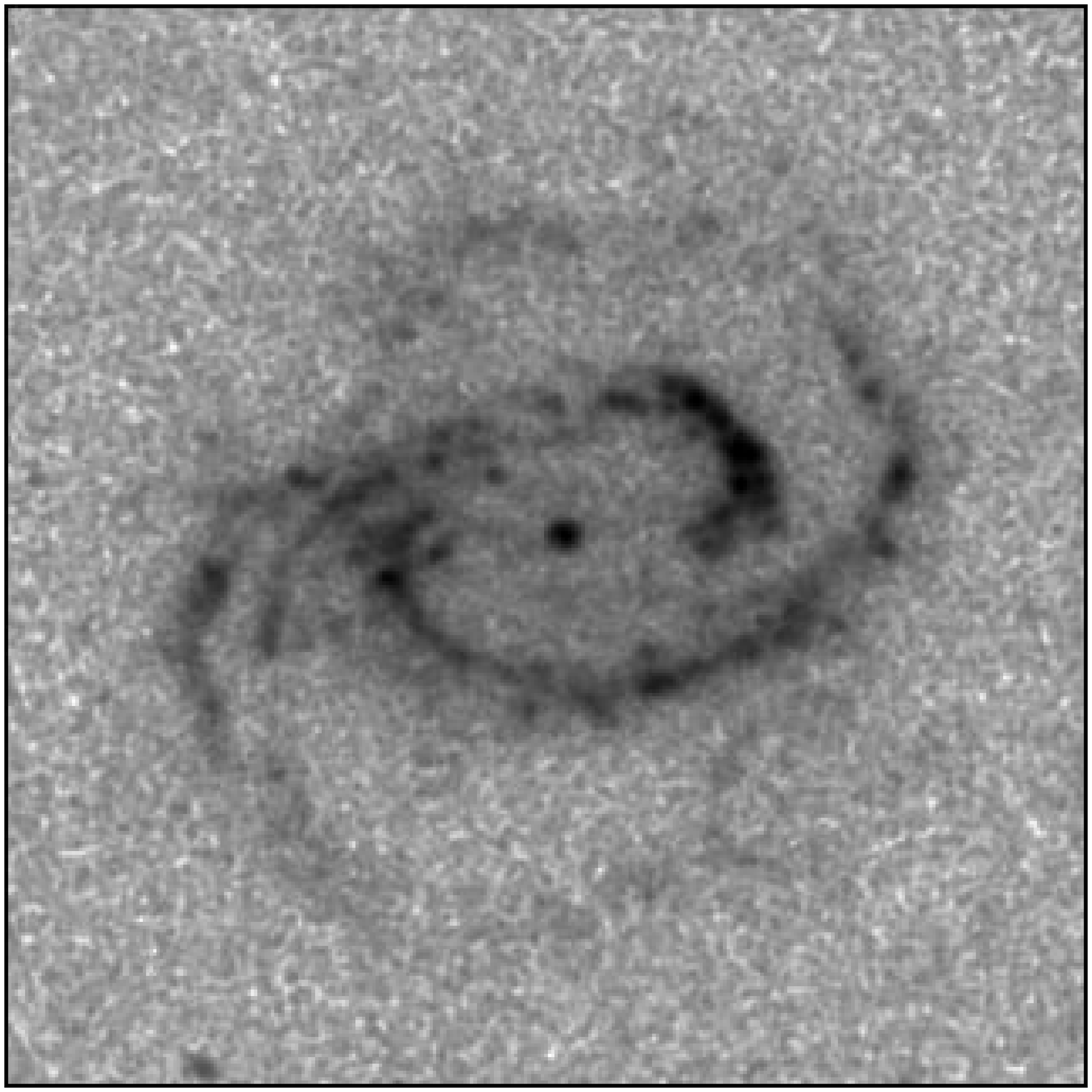} \\
\end{tabular}
\caption{Examples of the GALEX images, in the NUV band, for the
galaxies with pseudoring structures  ($R^{\prime}$), which
correspond to different types marked in the last column of
Table~1. The left plot: the  ring subtype~1, unclosed (NGC\,986, the
field-of-view size is $5^{\prime}$); the right plot: the ring subtype~2, clumpy
(NGC\,1300, the field-of-view size is $8^{\prime}$).}\label{fig4:Kostyuk_n}
\end{figure}

Table~2 summarizes the statistics of ring-shaped galaxies from the
list dividing by the morphological type and by the presence or absence of the  
UV radiation in the ring. Our list includes only 24 galaxies without bars
(SA type, see the second column of Table~1) and 94 barred galaxies
(of SB and SAB types). Although according to
Table~2 the number of galaxies with bars and rings is four times
larger than the galaxies with rings but without bars, we note
that the presence of a bar in the galaxy does not affect how often
the UV radiation is detected in the ring of a galaxy,
i.e., how often the current star formation proceeds in the rings. So in
the following discussion, we do not make a separation between the galaxies
with and without a bar.

\begin{figure}[tbp!!!]
\includegraphics[width=0.95\columnwidth]{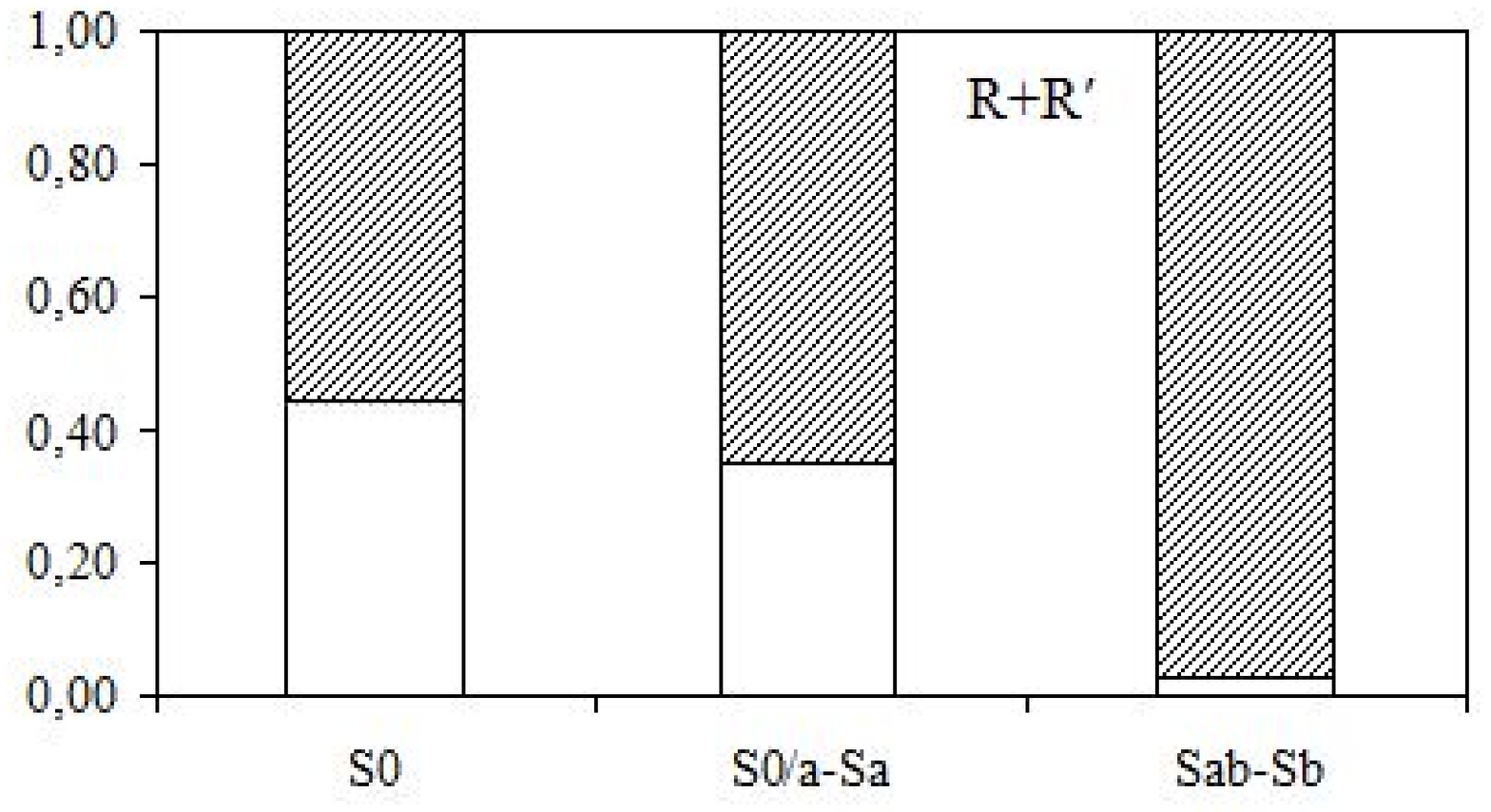}
\caption{Fraction of galaxies with ultraviolet radiation in the
ring (shaded) for different morphological types of all ring,
R$+$R$^{\prime}$, galaxies. The total number of
R$+$R$^{\prime}$ galaxies in each morphological type is
normalized to unity. }\label{fig5:Kostyuk_n}
\end{figure}

\begin{figure}[tbp!!!]
 \vspace{-3.5mm}
\includegraphics[width=0.95\columnwidth]{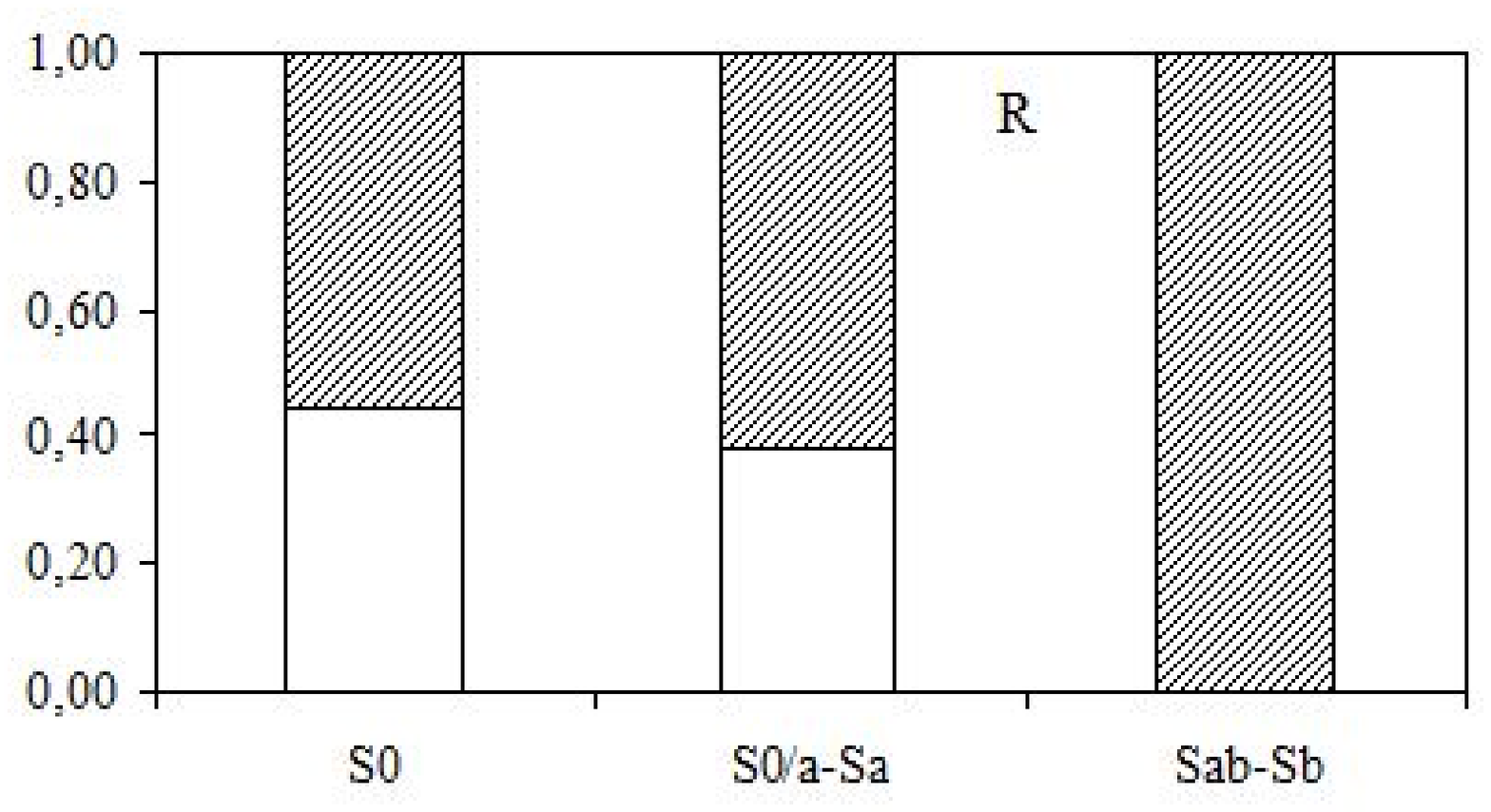}
\caption{Fraction of galaxies with ultraviolet radiation in the
ring (shaded) for different morphological types among the R
galaxies. The total
 number of galaxies in each morphological type is normalized to
unity. }\label{fig6:Kostyuk_n}
\end{figure}

\begin{figure}[tbp!!!]
 \vspace{-3.5mm}
\includegraphics[width=0.95\columnwidth]{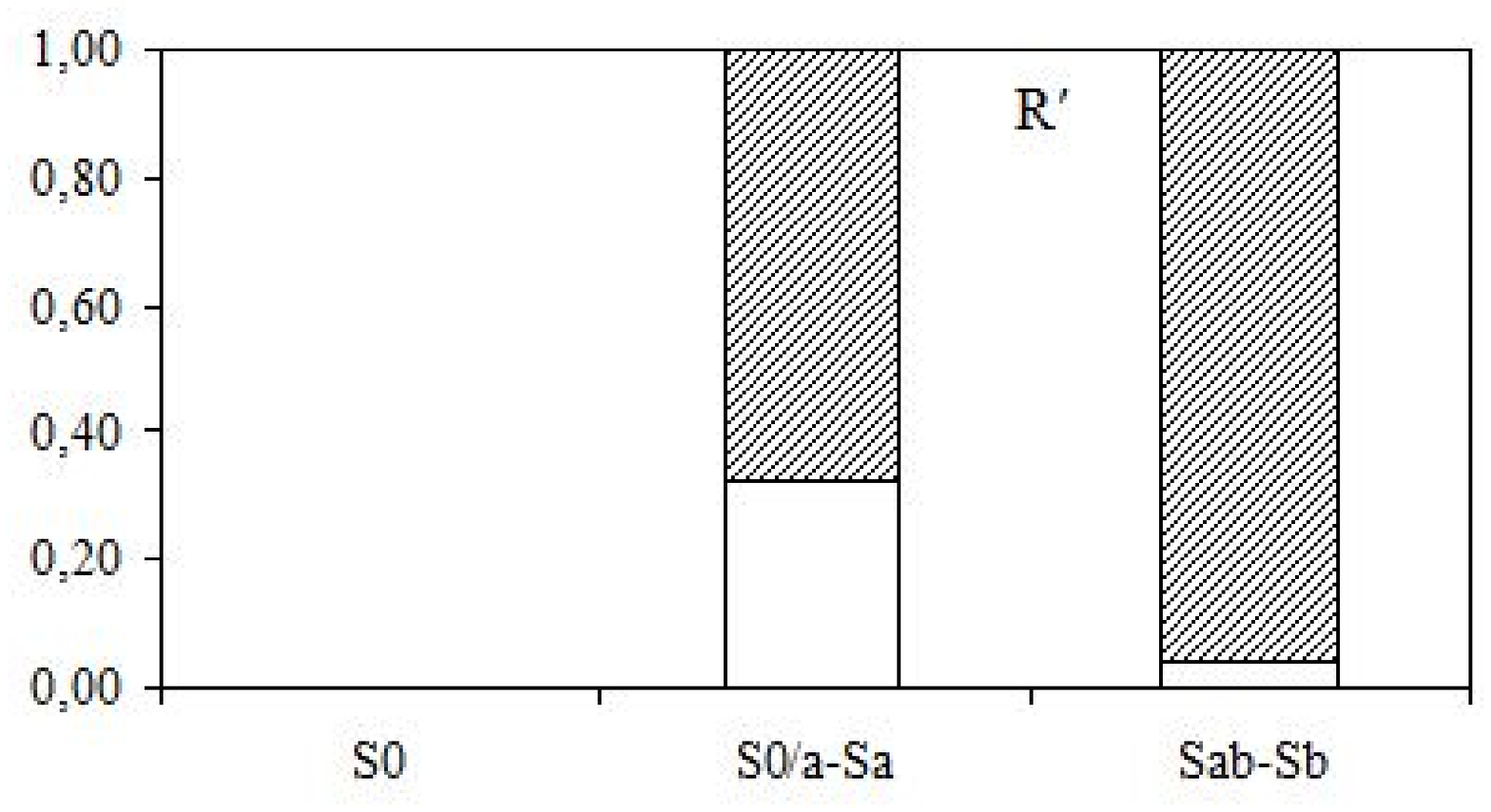}
\caption{Fraction of galaxies with ultraviolet radiation in the
ring (shaded) for different morphological types among the
pseudoring (R$^{\prime}$) galaxies. The total number of R$^{\prime}$
galaxies in each morphological type is normalized to unity.
The S0~galaxies never possess pseudorings.}\label{fig7:Kostyuk_n}
\end{figure}

Figures~5--7 present the histograms at which the fractions of the
galaxies with UV radiation in the ring (the shaded portion of the
column) among all R$+$R$^{\prime}$ ring-shaped galaxies are presented, 
and also those are shown separately for the ring~R and pseudoring R$^{\prime}$ galaxies
dividing according to their morphological types. All types of ring-shaped
structures are characterized by an increase in this fraction along the Hubble fork,
from S0 to Sb galaxies. But even at the minimum, for the S0 galaxies, it
stays at 56\%, which is about a half of all rings. Almost all the
spiral \mbox{Sab--Sb} galaxies (29 of 30) possess UV radiation in
their ring-shaped feature.

The average ratio of the major axis of the ring $D_r$ (the third
column of Table~1) to the major axis
of the outer isophote of the galactic disk  according to the RC3
data (the fourth column of  Table~1) is $0.81$  with a dispersion
of about $0.25$. Seventy percent of the galaxies in the list fall within
the ratio interval of \mbox{$0.6$--$1.0$}. Among the 
galaxies with the UV radiation in the ring-shaped structures
(84~galaxies), the subtype~2, which marks a clumpy structure (see the eighth column of Table~1), 
is the most frequent -- 61~galaxies. Interestingly,
this is true for all the morphological types of the galaxies. The
mean value of the  $k$ coefficient (the seventh column in
Table~1) is near~$6$, and it increases generally from S0 to Sb. Among the
ultraviolet-ring subtypes, mentioned in the eighth column of Table~1,
clumpy rings, subtype~2, have the mean $k$ coefficient almost two times 
larger than the other subtypes.

\section{DISCUSSION}

We have investigated the frequency of ongoing star
formation in the outer stellar rings of nearby early-type disk galaxies by
inspecting a representative sample (more than 100 objects)
taken from the \mbox{ARRAKIS} catalog~\cite{arrakis:Kostyuk_n}
compiling a list of galaxies with the rings seen at 4~$\mu$m. 
We have found that the regular outer stellar rings of the S0 galaxies  
contain young stars in about half the cases, while the pseudorings of 
spiral galaxies contain them almost always.

\renewcommand{\baselinestretch}{0.95}

\begin{table*}
\caption{Distribution of ring galaxies by the type of the
ring-shaped feature (R are rings, R$^{\prime}$ are pseudorings)
and the galaxy morphological type}
\medskip
\begin{tabular}{l|c|c|c|rr@{$\,\pm\,$}r|rr@{$\,\pm\,$}r|rr@{$\,\pm\,$}r}
\hline
\qquad Type & \multicolumn{3}{c|}{All} & \multicolumn{9}{c}{Among them, those having ``$+$'' in the UV column} \\
\cline{2-13}
 & R\,+\,R$^{\prime}$ & ~~~~R~~~~ & ~~~~R$^{\prime}$~~~~ & \multicolumn{3}{c|}{R\,+\,R$^{\prime}$} & \multicolumn{3}{c|}{R} & \multicolumn{3}{c}{R$^{\prime}$} \\
\hline
S0             & 25  & 25 & 0  & 14& (56\% & 10\%) & 14& (56\% & 10\%) &  \multicolumn{3}{c}{0} \\
of them, SB & 19  & 19 & 0  & 11& (58\% & 11\%) & 11& (58\% & 11\%) &  \multicolumn{3}{c}{0} \\
of them, SA & 6   & 6  & 0  & 3 & (50\% & 20\%) &  3& (50\% & 20\%) &  \multicolumn{3}{c}{0} \\
\hline
S0/a-Sa        & 63  & 29 & 34 & 41& (65\% &  6\%) & 18& (62\% &  9\%) & 23 &(68\% &  8\%) \\
of them, SB & 49  & 23 & 26 & 32& (65\% &  7\%) & 13& (56\% & 10\%) & 19 &(73\% &  9\%) \\
of them, SA & 14  & 6  & 8  & 9 & (64\% & 13\%) &  5& (83\% & 15\%) &  4 &(50\% & 18\%) \\
\hline
Sab--Sb        & 30  & 3  & 27 & 29& (97\% &  3\%) &  3& \multicolumn{2}{c|}{(100\%)}           & 26 &(96\% &  4\%) \\
of them, SB & 26  & 1  & 25 & 25& (96\% &  4\%) &  1& \multicolumn{2}{c|}{(100\%)}           & 24 &(96\% &  4\%) \\
of them, SA & 4   & 2  & 2  & 4 & \multicolumn{2}{c|}{(100\%)}           &  2& \multicolumn{2}{c|}{(100\%)}           &  2 &\multicolumn{2}{c}{(100\%)}           \\
\hline
All            & 118 & 57 & 61 & 84& (71\% &  4\%) & 35& (61\% &  6\%) & 49 &(80\% &  5\%) \\
of them, SB & 94  & 43 & 51 & 68& (72\% &  5\%) & 25& (58\% &  6\%) & 43 &(84\% &  5\%) \\
of them, SA & 24  & 14 & 10 & 16& (67\% & 10\%) & 10& (71\% & 12\%) &  6 &(60\% & 16\%) \\
\hline
\end{tabular}
\end{table*}

\renewcommand{\baselinestretch}{1.0}

Although by definition of their morphological type lenticular galaxies are thought  
to be devoid of large-scale star formation in the disks, in fact, at a  closer look, 
is not exactly so. In  21\%\ of all S0 galaxies the ultraviolet space telescope GALEX 
observes an extended signal, that indicates extended regions of current star
formation~\cite{salim:Kostyuk_n}. The distribution of this extended
star formation is quite curious. Recently,  Salim et~al.~\cite{salim:Kostyuk_n} 
examined a sample of early-type galaxies known to have ultraviolet excess. After 
observing them at the Hubble Space Telescope and constructing well-resolved UV images,  
the kind of morphology of the UV images was found to be related with the galaxy
morphological type: in six ellipticals star formation is concentrated in a
small (unresolved) area, whereas in 15 of 17 S0 galaxies extended star
formation was noticed, and in all 15 cases, the star formation
morphology is ring-like. The rings may be of various sizes: for
the narrow rings of star formation, the authors of~\cite{salim:Kostyuk_n} 
obtained a mean radius of $6.5$~kpc, for the wide rings---\mbox{$16$--$20$ kpc.} 
There still exists a rare morphological subtype of `a disk with a hole'; however, 
these are actually rings with large outer radii. Among the S0 and Sa
early-type disk galaxies possessing rings of star formation in 
the sample~\cite{salim:Kostyuk_n}, there are equal numbers of galaxies 
with and without bars~(8~and 11~objects respectively), so the resonance 
nature of the most of these rings is not quite obvious. We would like to  
note that the conclusion about the dominance of the ring-like morphologies 
as concerning the distribution of starforming regions in lenticular galaxies
has already been reported earlier. As early as in 1993, during deep searching 
for H$\alpha$ emission in the disks of lenticular galaxies rich in neutral hydrogen,
Pogge and Eskridge~\cite{pogge_esk:Kostyuk_n} found that star
formation could be detected in half the cases, and that it was always
organized as rings. It was a surprise to find that the presence or absence of 
star formation was not related to the amount of fuel for star formation 
(to the amount of neutral hydrogen). Pogge and Eskridge~\cite{pogge_esk:Kostyuk_n} 
concluded then that star formation in S0s was probably triggered by some external 
kinematical factor, quite different from the gravitational instabilities,  
controlling star formation in thin large-scale disks of late-type spiral galaxies. 
Some claims about finding star formation in outer rings,
where the gaseous component should in theory
be stable against the processes of fragmentation, came from the
detailed study of the outer neutral-hydrogen rings in early-type
spiral galaxies by \cite{noor05:Kostyuk_n}. The most natural
additional mechanism which may provoke star formation ignition 
in the gaseous disks with the {\it mean} surface density under the Kennicutt 
threshold~\cite{kenni_sf:Kostyuk_n} seems to involve shock waves. This
leads us to the suggestion of the cold gas accretion from 
outside, perhaps from satellites on inclined orbits (to provide a compression
shock wave in the main galactic disk due to vertical impact) as the dominant
mechanism to form regular outer rings of star formation in the disks of early-type 
galaxies, where surface density of the gas, as known (see,
for example~\cite{oo99:Kostyuk_n,oo10:Kostyuk_n}), is significantly smaller
than that in the disks of late-type galaxies, and moreover, as a rule is
insufficient to support star formation on its own. At the same time, the
distribution of neutral hydrogen in early-type galaxies (where
the neutral hydrogen is found) is much more extended than that
in spiral galaxies: regular H\,I structures in S0s can reach up to
200~kpc in diameter~\cite{oo07:Kostyuk_n}. Basing on multi-colour
surface photometry, Afanasiev and Kostyuk~\cite{af_kost:Kostyuk_n} have
shown that the ring-shaped galaxies from the list of~\cite{kostuk75:Kostyuk_n}
belong mostly to the early morphological types, in agreement with the conclusions 
of Comer\'on et~al.~\cite{arrakis:Kostyuk_n}. However,
in~\cite{af_kost:Kostyuk_n} it was also noticed that the galaxies
with outer rings have in average more extended stellar disks than 
the galaxies without rings. It is a strong indication on the further
build-up of the outer parts of the disks through stimulated star
formation in the outer rings as a result of cold gas accretion from
outside.

It seems possible that the outer pseudorings of early-type spiral
galaxies, where star formation proceeds in almost 100\%\ of the
cases, have the same genesis as the phenomenon of the so-called
XUV disks (eXtended UltraViolet disks~\cite{xuv1:Kostyuk_n,lemonias:Kostyuk_n}), 
observed in approximately 20--30\%\ of all disk galaxies, from early to 
late types, in the nearby Universe. Broken (unclosed) pseudorings may well 
be tightly wound spiral density waves, spreading outward into extended gas
disks from the inner regions of the galactic disks. These generate also
shock waves capable to stimulate star formation in diffuse gaseous 
medium~\mbox{\cite{bush08:Kostyuk_n,bush10:Kostyuk_n}}. However,
the origin of the XUV disks, which can be equally often found both in
low-luminosity starforming galaxies and in massive galaxies that inhabit
the   ``red sequence'' and ``green valley''~\cite{moffett:Kostyuk_n}, 
is presently also attributed to recent events of outer cold gas 
accretion~\cite{lemonias:Kostyuk_n,stewart:Kostyuk_n,holwerda:Kostyuk_n}.
Hence, for pseudorings also, the external origin 
through the accretion of cold gas with a high angular
momentum onto the periphery of a galactic disk remains to be
the most attractive. Perhaps for these structures we would prefer
a slow, smooth accretion close to the  galactic disk symmetry plane.

\section{Acknowledgements}
We have made use of the data from the NASA/IPAC
Extragalactic Database (NED) and the Spitzer Space Telescope which
are operated by the Jet Propulsion Laboratory, California
Institute of Technology, under contract with NASA. Some data (NASA
GALEX mission) were obtained from the Mikulski Archive for Space
Telescopes (MAST), which is supported by the NASA Office of Space
Science via grant NNX09AF08G and by other grants and contracts.
This study has been performed at the expense of the Russian
Science Foundation (project \mbox{No.~14-22-00041}).

\end{document}